\documentclass[pre,aps,twocolumn]{revtex4}
\usepackage[latin1]{inputenc}
\usepackage[spanish]{babel}
\usepackage{graphicx}

\begin{document}

\title{Functional gradients through the cortex, multisensory integration \\ and scaling laws in brain
dynamics}
\author{Isabel Gonzalo-Fonrodona }

\affiliation{Departamento de \'Optica. Facultad de Ciencias
F\'\i{}sicas.
\\ Universidad Complutense de Madrid. Ciudad Universitaria s/n.
28040-Madrid. Spain\\ E-mail: igonzalo@fis.ucm.es}

\begin{abstract}

In the context of the increasing number of works on multisensory
and cross-modal effects in cerebral processing, a review is made
on  the functional model of human brain proposed by Justo Gonzalo
(1910-1986), in relation to what he called central syndrome
(caused by unilateral lesion in the parieto-occipital cortex,
equidistant from the visual, tactile and auditory projection
areas). The syndrome is featured by a bilateral, symmetric and
multisensory involvement, and by a functional depression with
dynamic effects dependent on the neural mass lost and related to
physiological laws of nervous excitability. Inverted or tilted
vision as well as tactile and auditive inversion, under minimum
stimulus, appears as a stage of incomplete integration, being
almost corrected under higher stimulus or facilitation by
multisensory integration. The syndrome reveals aspects of the
brain dynamics that suggest  a functional continuity and unity of
the cortex. A functional gradients scheme was proposed in which
the specificity of the cortex is distributed with a continuous
variation. This syndrome is interpreted as a scale reduction in
the nervous excitability of the system, the different sensory
qualities being affected allometricaly according to scaling laws.
A continuity from lower to higher sensory functions was proposed.
The sensory growth by an increase of the stimulus or by
multisensory facilitation is found to follow approximately power
laws, that would  reflect  basic laws of biological neural
networks. We restrict the analysis to the visual
system. \\


\end{abstract}
%\end{frontmatter}
\maketitle

\section{Introduction}
In recent years, many works reported on multisensory and
cross-modal effects in integrative cerebral processes (e.g.,
\cite{Martuzzi07,Gillmeister07,Kayser07,Kayser072,Alvarado07,Diederich07,Poggel06,Bizley06,
Frassinetti05,Macaluso05,Pascual04,Calvert04} referring only to
some recent works among many others). It was found that
cross-modal interactions can affect activity in cortical regions
traditionally regarded as ``unimodal", as it occurs for example in
the contribution of visual cortex to tactile perception
\cite{Sathian02,Pascual97,Pascual04}. Some of these multisensory
interactions have been  revealed by functional magnetic resonance
imaging, positron emission tomography and analysis of blood oxygen
level. According to some authors, all these findings suggest that
longstanding notions of cortical organization need to be revised
to include multisensory interactions as a inherent component of
functional brain organization \cite{Martuzzi07} (see also
\cite{Wallace04} and references therein).

In this order of ideas, we recall in the present work the
pioneering research of Justo Gonzalo
(1910-1986)\cite{Gonzalo45,Gonzalo51,Gonzalo52}. This author
studied about one  hundred selected patients with brain injuries
from the Spanish Civil War (1936-39). Twelve of them with brain
injury in the parieto-occipital zone and special features led to
characterize what the author called ``central" syndrome of the
cortex, and to propose a dynamic interpretation of cerebral
localizations. This allowed the author to explain the cases of
first hand and others reported in the literature as the
much-discussed Schn. case \cite{Goldstein18}. The author developed
a functional model of some aspects of brain dynamics based on
functional gradients through the cortex and scaling laws of
dynamic systems, supported by physiological laws of nervous
excitability. It was exposed in part in \cite{Gonzalo45} and
further developed in
\cite{Gonzalo51,Gonzalo52,Gonzalo96,Gonzalo97,Gonzalo99,Gonzalo01,Gonzalo03,Gonzalo07,Gonzalo07P}.
It offers a framework to understand at a functional level, several
aspects of the integrative cerebral process. The immediate
repercussion of this research (e.g.,
\cite{Critchley53,Bender48,Ajuriaguerra49}) diverted later towards
the development of cerebral processing models
\cite{Delgado78,Mira87,Mira95,Manjarres00,Gonzalo01,Gonzalo03},
apart of historical notes \cite{Arias04,Barraquer05}.

The central syndrome \cite{Gonzalo45,Gonzalo51,Gonzalo52} is
originated from a unilateral parieto-occipital lesion equidistant
from the visual, tactile and auditory projection areas (``central"
$\,$ zone) as shown in Fig.1, A (the middle of area 19, the
anterior part of area 18 and the most posterior of area 39, in
Brodmann terminology). It presents a multisensory and symmetric
affection; visual, tactile and auditory systems are equally
affected, in all its functions and with symmetric bilaterality. In
the visual system for example, in addition to other disorders,
there is a bilateral and symmetric concentric reduction of the
visual field (Fig. \ref{CEREBRO}, A) with gradation of the
involvement from the center to the periphery. This syndrome
involves a loss of rather unspecific (or multispecific) neural
mass and is characterized by a functional {\em depression} shown
for example in the excitation threshold curves in Fig.
\ref{EXCITATION} \cite{Gonzalo45,Gonzalo52}. In this figure we see
that the thresholds of the acute case M are greater than those of
the less intense case T (with a smaller lesion), and the
thresholds of T are greater than in a normal man. This depression
presents dynamic phenomena related to transformations of the
central nervous excitability which depend on the quantity of
neural mass lost. These phenomena are \cite{Gonzalo45,Gonzalo52}:

\begin{figure}
\begin{center}
\includegraphics[width=4cm]{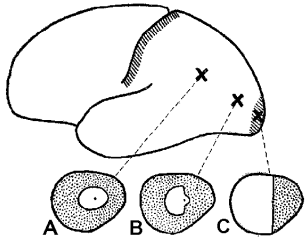}
\caption{\label{CEREBRO}Position of cortical lesions and
respective visual fields (involved zones are dark). A: central, B:
paracentral, C: marginal or peripheral, syndromes. (Adapted from
Fig. 1(a) of \cite{Gonzalo96} with permission of the MIT Press).}
\end{center}
\end{figure}

\begin{figure}
\begin{center}
\includegraphics[width=6cm]{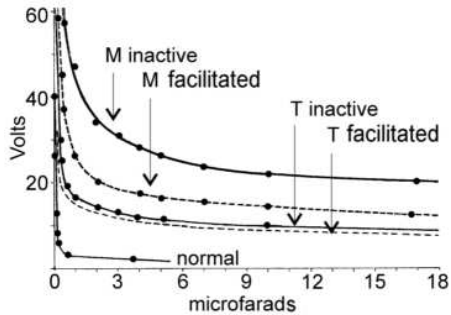}
\caption{\label{EXCITATION}Excitation threshold curves for
electrical stimulation of the retina (cathode on eyelid) in acute
central syndrome case M, less intense case T, each one in inactive
state and facilitated state by strong muscular stress (see the
text), and in a normal case. Electrical intensity (indirectly
given by volts) versus time (given by microfarads) necessary to
obtain minimum luminous sensation. (From Fig. 2 of
\cite{Gonzalo07} with permission of Viguera Editores, S.L.).}
\end{center}
\end{figure}

\begin{figure}
\begin{center}
\includegraphics[width=3.5cm]{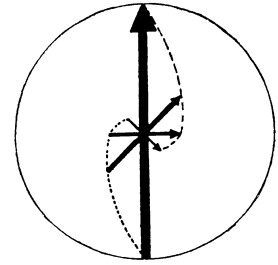}
\caption{\label{FLECHA}Perceived tilt and diminution in size of a
vertical test arrow in the center of the visual field of right eye
in central syndrome, as the illumination of the arrow diminishes.
Spiral branches described by the extremes of the arrow are
indicated. (Adapted from Fig. 3(a) of \cite{Gonzalo96} with
permission of Springer Science and Business Media).}
\end{center}
\end{figure}

(a) Functional disgregation or dissociation of sensory qualities
(normally united in perception) according to their excitability
demands, i.e., sensory qualities are gradually lost as the
stimulus intensity diminishes. For instance, when the illumination
of an upright white arrow was diminishing, the arrow perceived is
at first upright and well defined, next it is perceived to be more
and more rotated in the frontal plane (Fig. \ref{FLECHA}) becoming
at the same time smaller and losing its form and colors in a
well-defined physiological order, the perceived tilt (almost
inverted vision in the acute syndrome case M) being dependent on
the size, distance, illumination and exposure time of the test
object. In the tactile system, five stages  were distinguished
successively in the dissociated perception of a mechanical
pressure stimulus on one hand, as the energy of the stimulus was
increased: first, primitive tactile sensation without
localization, then, contralateral localization (inverted
perception), and finally, normal localization which required
intense stimulus. A mobile stimulus was perceived in the inversion
phase with a very shortened trajectory (approximately 1/10 in case
M). In the auditory system, there was a dissociation between
simple sonorousness (weak stimulus) and real tone (stronger
stimulus) of a particular sound. Contralateral localization
(spatial inversion) of a sound stimulus occurred only in the most
acute case M when the stimulus was weak and the patient was in
inactive state (free of facilitation). The inverted perception
always lacked tonal quality. In language, diverse aphasic aspects
occurred depending on the stimulation, it will be exposed in a
future work.

(b) Capability to multisensory facilitation (the perception of a
stimulus is improved by adding simultaneously one or more
different stimuli). It modifies the cerebral system essentially,
becoming more rapid and excitable, i.e., it supplies in part the
neural mass lost in the ``central" lesion, thus reducing the
mentioned dissociation. It was found that a strong muscular stress
was very efficient at improving the perception (see the
facilitated cases in Fig. \ref{EXCITATION}. Other types of
facilitation to visual perception were binocular summation and
tactile and acoustic stimuli, for example.

(c) Capability to temporal summation, which is merely a particular
means of stimulation. The slowness of the cerebral system in
central syndrome makes the cerebral excitation to a short stimulus
to decay slowly. If a second stimulus arrives before this
excitation has completely fallen down, excitations are summed up,
so that it is possible to achieve an excitation threshold to
produce a sensory perception, reducing so the mentioned
pathological dissociation. It was shown \cite{Gonzalo01} that it
is possible to model the temporal dynamics of simple sensory
functions as a first-order linear time-dispersive system having as
a first approach a time response $\chi_0 e^{-a/t}$ ($t$ is time)
to a short impulse stimulus. The reaction velocity $a$ and the
permeability to the excitation $\chi_0$ of the cerebral system
decrease when the deficit (due to the lesion) of active ``central"
neural mass is greater, but the ratio $a/\chi_0$ increases
significantly and can be considered as a constant for each sensory
function of a given cerebral system without facilitation.

Thus, the importance of  the central syndrome (say, symmetric
multisensory syndrome) lies in the fact that changes in the
stimulation intensity (and then in nervous excitation) reveal
aspects of the dynamics of the cerebral cortex,  multisensory
integration, and organization of sensory structures combined with
the direction function manifested in the tilted or almost inverted
perception. Spatial inversion appears as an essential fact in the
sensory organization. In a recent review \cite{Gonzalo07}, the
inverted or tilted perception disorder observed and interpreted by
Gonzalo, is put in relation with the increasing number of reported
cases with this anomaly in the last years (for example
\cite{Arias01,Arjona02,Malis03,Hernandez06,Unal06,Kasten06} among
many others  reviewed in \cite{Gonzalo07}). In tilted vision, a
rather similar degradation of the perception to that referred in
(a) was reported in \cite{Kasten06}.

We expose in the following the  principal features of the model
proposed by Gonzalo highlighting the connection with recent
research.

\section{Functional cerebral gradients}

From  the first hand cases  and  others reported in the
literature, the diverse syndromes were ordered according to the
position and magnitude of the lesion \cite{Gonzalo52}. The central
syndrome refers, as said above, to lesions in the ``central" zone,
equidistant from the visual, tactile and auditive projection
paths. Syndromes corresponding to lesions in the projection paths
were called in this context ``marginal" $\,$ or ``peripheral"
syndromes, the defect -functional suppression- is restricted to
the contralateral half of the corresponding sensory system and
scarcely present dynamic effects. Intermediated syndromes were
called ``paracentral" syndromes, with different degree in the
bilateral symmetry involvement (see Fig. \ref{CEREBRO}).

\begin{figure}
\begin{center}
\includegraphics[width=7cm]{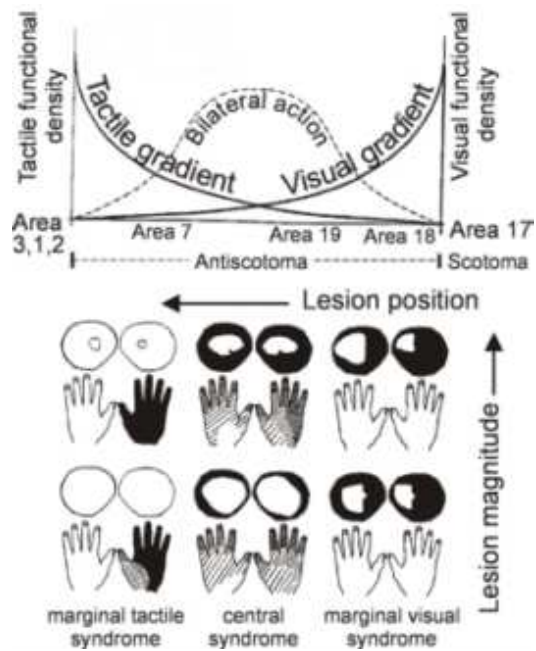}
\end{center}
\caption{\label{GRADIENTES}Lower part: visual fields and tactile
sensitivity of cases ordered according to the position and
magnitude of the lesion. The degree of the defect is greater in
darker regions. Upper part: scheme of visual and tactile
functional densities $f$, and the unspecific functional gradation
which represents the multisensory and bilateral effect. Brodmann
areas are indicated. (Adapted from Fig. 5 of \cite{Gonzalo96} with
permission of The MIT Press).}
\end{figure}

The complete gradation found between the central syndrome, and a
marginal syndrome, through the variety of paracentral syndromes
lead to the definition of two types of continuous functions
through the cortex called ``cerebral gradients" \cite{Gonzalo52},
shown schematically in Fig. \ref{GRADIENTES}. One type comprises
the specific sensory functional densities, of contralateral
character, with maximum value in the respective projection area
and decreasing gradually towards more ``central" zone and beyond
so that the final decline of the specific visual function density,
for example, must reach all the tactile area, as shown in Fig.
\ref{GRADIENTES}, as well as other specific areas including their
primary zones. This type of function takes into account and
combines the factors of position and magnitude of the lesion,
since the more ``central" is the lesion, the more extensive must
be the lesion to originate the same intensity in a specific
anomaly. For a given position of the lesion, its magnitude
determines de degree of functional depression. The other type of
function, of unspecific (or multispecific) character, is maximum
in the ``central"  $\,$ region (where the decline of the above
mentioned specific functions overlap) and vanishes towards the
projection areas. It represents the multisensory effect in the
anomalies and the bilaterality or interhemispheric effect by the
action of the corpus callosum. Each point of the cortex is then
characterized by a combination of specific contralateral action
with unspecific ``central" $\,$ and bilateral action.

In the visual system for example, for the visual function to be
normal, the action of the region with greatest visual sensory
function density is not enough, and the whole specific functional
density, say f, in gradation through the cortex, must be involved
in the integrative cerebral process, leading to the normal sensory
visual function $F$. Analogously for the other senses and
qualities. In a schematic way, $f$ is related to the derivative of
this integration through the cortex, which justifies it was named
gradient. This function $f$ takes into account the density of
specific neurons through the cortex and their connections,
representing the dynamic aspect of its anatomic basis. A sensory
signal in a projection area is only an inverted and constricted
outline that must be elaborated (magnified, reinverted, ...),
i.e., integrated over the whole region of the cortex where the
corresponding specific sensory functional density $f$ is extended.
Magnification would be due to the increase in recruited cerebral
mass, and reinversion due to some effect of cerebral plasticity,
(following an spiral growth as the growth of the arrow in Fig.
\ref{FLECHA}). In the visual system, reinversion and
bilateralization would occur in the 18 and 19 Brodmann areas where
the sensory representation is already reinverted. A remarkable
result is, for example, the significant participation of the
traditionally  ``extravisual"  $\,$  cortex in the maintenance of
the visual field.

\begin{figure}
\begin{center}
\includegraphics[width=4.6cm]{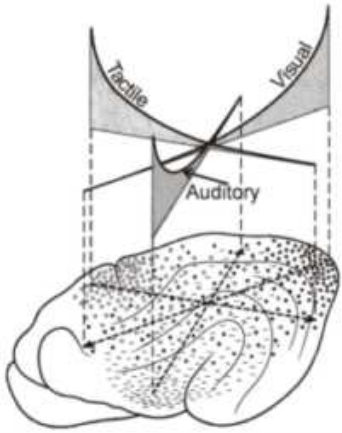}
\end{center}
\caption{\label{3GRADIENTES}Schematic visual, tactile and auditory
functional densities.}
\end{figure}

The projection zones where the respective specific functional
gradients are maximum, are highly organized and differentiated (in
biological terms), i.e., highly specialized. They are
phylogenetically the oldest zones of the isocortex and the nervous
activity has an anatomic representation. In contrast, the
``central" zone which is more recent, is rather unspecific with
capacity for adaptation or learning. It is very small in animals,
even in other mammals, but it has great extension in man. A
schematic representation of the visual, tactile and auditory
gradients is shown in Fig. \ref{3GRADIENTES}.

Certain  sensory functional densities would arise from the
superposition of functional gradients of different systems, as in
the case of primary alexia where the corresponding functional
specific (lexical) gradient would result from the superposition of
auditory and visual ones, leading to a lexical bell gradient
between the two systems. In some cases, a gradient with
hemispherical dominance has to be added.

As opposed to the rigid separation of zones, the functional
gradients  account for a functional continuity and physiological
heterogeneity of the cortex, this one being subjected to a common
principle of organization. This scheme, mere abstraction of the
observed facts, offers a dynamic conception of quantitative
localizations which permits an ordering and an interpretation of
multiple phenomena and syndromes. Also, this scheme is in relation
to recent works
 suggesting that traditional specific cortical
domains are separated from one another by transitional
multisensory zones \cite{Wallace04}, and that multisensory
interactions occur even in the primary sensory cortices
\cite{Martuzzi07,Kayser07,Sathian02}.

\section{Scale reduction in central syndrome. Allometric scaling laws in the display of sensory qualities}

\begin{figure}
\begin{center}
\includegraphics[width=6cm]{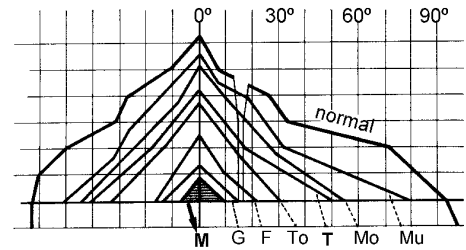}
%OJO que he puesto figura en vez de figure !!!!
\end{center}
\caption{\label{SENSIBILITY}Sensibility profiles of visual fields
in different central syndrome cases (M and T included) and in a
normal case. (Adapted from Fig. 2 of \cite{Gonzalo97} with
permission of Springer Science and Business Media). }
\end{figure}

The cerebral system after a ``central" lesion, once the new
dynamic equilibrium is reached, was considered  a scale reduction
in the nervous excitability of the cerebral system since the
originated functional depression maintains, nevertheless, the same
cerebral organization as in normal case. This can be appreciated,
for example, in the hypoexcitability of the nervous centers shown
in the excitability (Fig. \ref{EXCITATION}) and luminosity
threshold curves, and also in the concentric reduction of the
visual field, its sensibility profile (see Fig. \ref{SENSIBILITY})
and visual acuity. All these functions experienced a down shifting
in their values, but keeping the same form as in a normal case
\cite{Gonzalo52}.

Then, the concept of dynamic similitude, according to which
different parts of a dynamic system change differently under a
change in the size of the system, was applied. In particular, in
biological growth, the sizes of two parts (say x and y) of a
biological system are approximately related  by a scaling power
law of the type
\begin{equation}
 y = A x^n ,     \label{POWER}
 \end{equation}
$n$ being different for different parts ($y_1, y_2$, ...) of the
system. These parts change then differently, i.e., allometrically
\cite{Perkkio85}. This power relation means that the rates of
growth of the two parts compared are proportional, i.e.,
$(1/y)(dy/dt)=n(1/x)(dx/dt) $ (obtained from Eq. (\ref{POWER}) by
taking the logarithm and differentiating). These ideas were
applied by  Gonzalo \cite{Gonzalo97} to the growth (or reduction)
of the sensory qualities, linked each one to a nervous excitation
demand (i.e., to a quantity of neural mass). An allometric
variation of the sensory qualities was then proposed.

\begin{figure}
\begin{center}
\includegraphics[width=6cm]{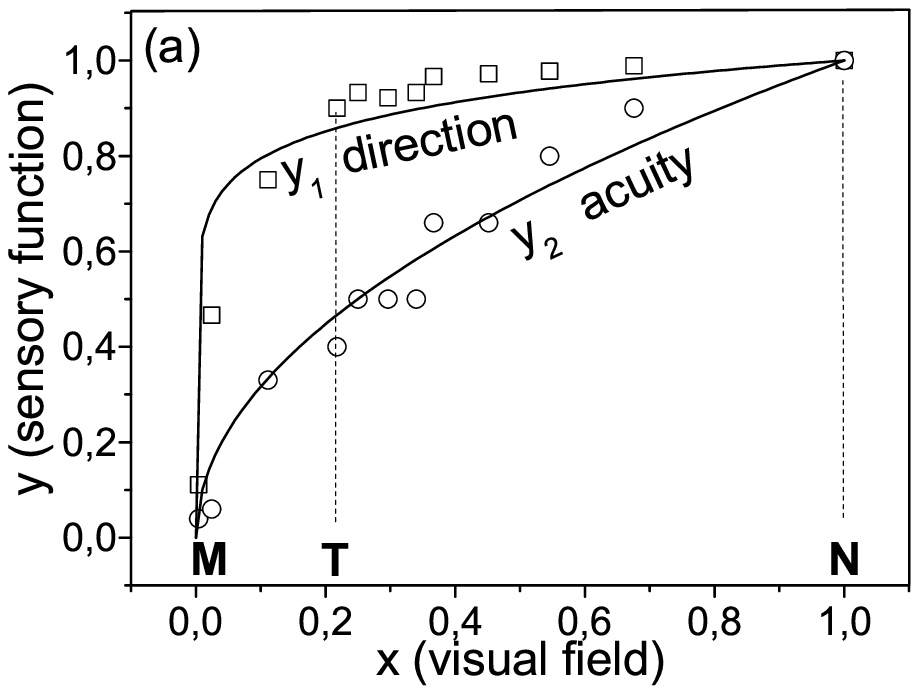}\\
\vspace*{0.3cm}
\includegraphics[width=6cm]{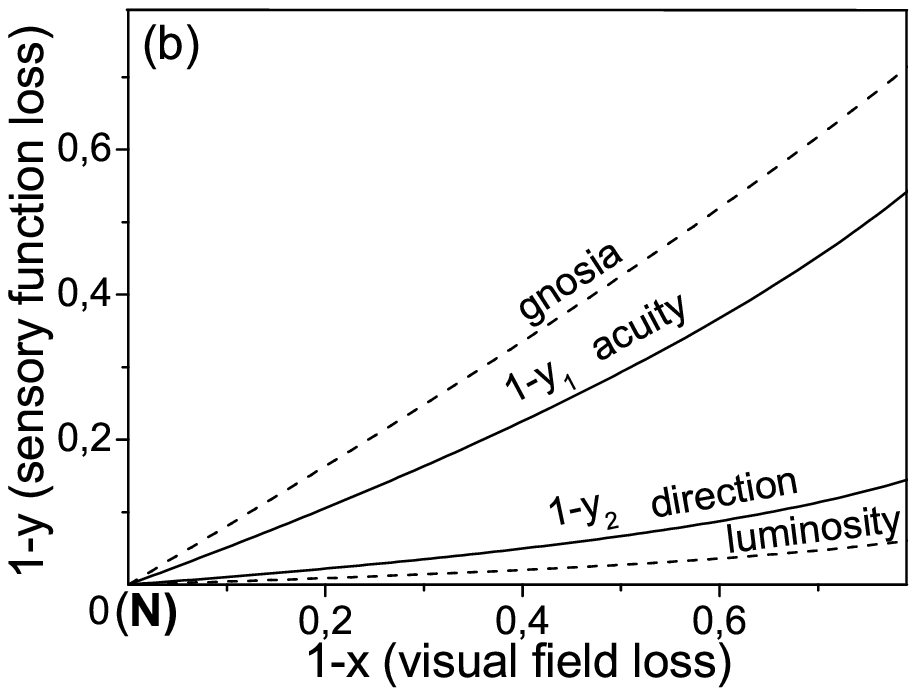}
\end{center}
\caption{\label{ALLOMETRIC} (a)Normalized direction function
(squares), fitting to $y_1= x^{0.5}$, and normalized acuity
function (circles), fitting to $y_2= x^{0.1}$, for central vision,
versus normalized untouched visual field surface $x$ of the
observing eye. Cases M, T and normal (N) are indicated. (b) Curves
for the loss of visual acuity, $1-y_1$, and loss of direction,
$1-y_2$, versus loss of the visual field, $1-x$, for the same
conditions and normalizations as in (a). Qualitative indications
for the loss of the higher sensory function gnosia and lower
sensory function luminosity are shown as examples.}
\end{figure}

For the pure central syndrome cases studied \cite{Gonzalo52},
normalized values of the visual direction function
 $y_1$ and of the
acuity function $y_2$, versus normalized untouched  visual field
surface $x$, are plotted in Fig. \ref{ALLOMETRIC} (a). The acute
case M (with considerable neural mass lost), the less intense case
T (with less neural mass lost than M) and a normal case N are
indicated. Errors are greater in cases with intense central
syndrome because of their high sensory degradation and the high
capability to facilitation by temporal and multisensory summation.
The values plotted are normalized with respect to the normal
value. The direction function is considered 0º for the total
inverted perception of the upright test arrow, and 180º for the
upright perception (normal), i.e., the maximum value is achieved
in a normal integrative process from the inverted signal in the
projection path. For a normal case, N, the normalized values $x$,
$y_1$ and $y_2$  are the unity, the maximum value. For case M, the
corresponding values are very small since the non-normalized
values are, 6 degrees for the visual field width, 0.04 for the
acuity and 160 degrees for the perceived direction.

We can see in Fig. \ref{ALLOMETRIC}(a) that the data fit
approximately to power laws. Since $y_1$ and $y_2$ are 1 for
$x=1$, the value $A$ of Eq. (\ref{POWER}) is 1 and the approximate
allometric power laws found are $y_1= x^{0.5}$ for the direction
function and $y_2= x^{0.1}$ for the acuity function. If the widths
of the visual fields are considered instead of the surface values
$x$, there is no evidence of power laws.

In order to appreciate the loss of the sensory functions in the
pathological disgregation of the  sensorium in central syndrome,
the loss of  direction function $(1-y_1)$ and the loss of  visual
acuity $(1-y_2)$ are plotted in Fig. \ref{ALLOMETRIC}(b) versus
the loss of visual field $(1-x)$, together with a qualitative
representation of the loss of other functions: an elementary one
as luminosity and a higher one as gnosia, according to the
observed facts. The origin of the graphs corresponds to a normal
man (N). We can see that for a particular loss of visual field
(due to the scale reduction originated by a particular central
lesion), a split of the different qualities occurs so that the
higher  ones (e.g., gnosia) are loss in a higher degree than the
lower (elementary) ones (e.g., luminosity). The order of the split
corresponds to the order of complexity (or excitability demand) of
the sensory functions and to the order they are lost due to the
shifts in their threshold excitabilities.

This is the formal description of the mentioned functional
depression where the most complex qualities, with greatest nervous
excitability (and integration) demand, become lost or delayed in
greater degree than the most simple ones (with lower excitability
demand). Sensations usually considered as elementary are then seen
to be decomposed into several functions, one of them being the
direction function, thus revelling up to a certain extent, the
organization of the sensorium. Very small differences in the
excitation of different qualities occur already in the normal
individual (in colors for example), and they grow considerably in
central syndrome. In this context, it could be said that the
cerebral system of the normal man works like an almost saturated
system, in the sense that a very low stimulus induces cerebral
excitation enough to perceive not only the simplest sensory
functions but also the most complex ones in a synchronized way.

In relation to the scheme of cerebral gradients, and for a sensory
system, the functional gradient in  normal man must be understood
as an ensemble of several functional densities $f_i$ for different
qualities in gradation through the cortex, each one with different
slope according to its excitability and integration demand. In the
new equilibrium achieved in central syndrome the deficit of neural
excitation affects these functions $f_i$ differently. Their
incomplete cerebral integration gives place to the respective
reduced sensory qualities $F_i$, each one reduced differently
(allometrically). The mentioned disgregation phenomenon of the
perception in the functional depression corresponds then  to
different stages of incomplete integration in different degree for
the different qualities.

\section{Sensory growth, multisensory effects and more scaling power laws }

In a reduced cerebral system as was described above, the sensory
level may grow by intensifying the stimulus (it includes iterative
temporal summation) and by sensory facilitation by adding other
different stimuli, as was said in the first section. These
mechanisms involve an increase of central nervous excitation being
able to compensate in part for the neural mass lost. This
capability to improve the perception is greater as the magnitude
of the central lesion is greater, and is null or very low in a
normal man. The best example is the extreme case M, who could
approach the physiological and sensory level of the less intense
case T, by the use of strong muscular stress (M facilitated) as
shown in Fig. \ref{EXCITATION}. The inactive state, free of
facilitation is difficult to achieve in acute cases as M since
mere postural changes or weak stimuli modify the perception.

\begin{figure}
\begin{center}
\includegraphics[width=5.5cm]{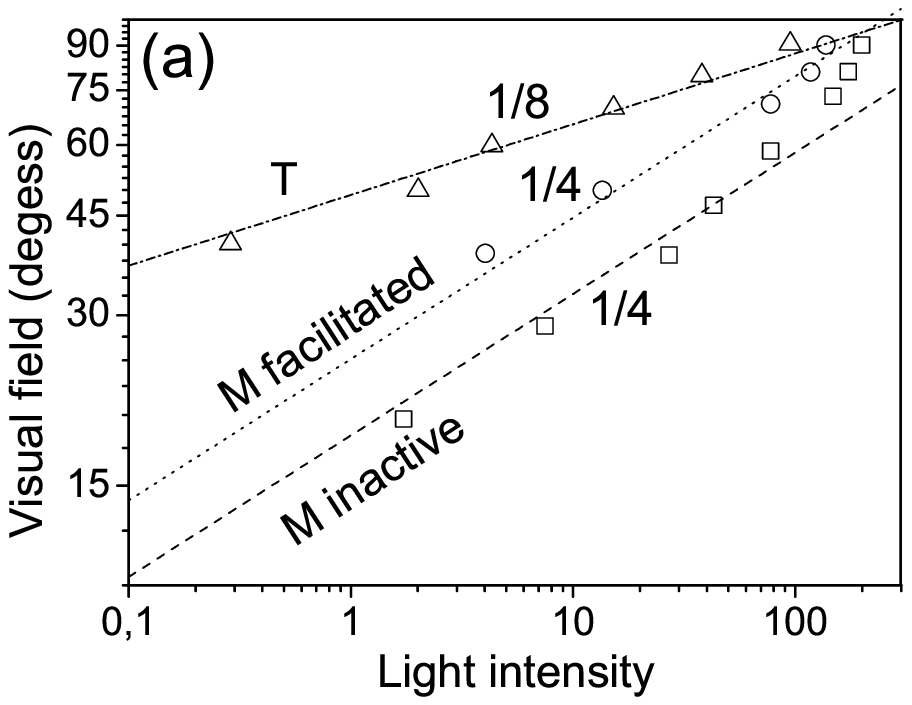}\\
\vspace*{0.3cm}
\includegraphics[width=5.5cm]{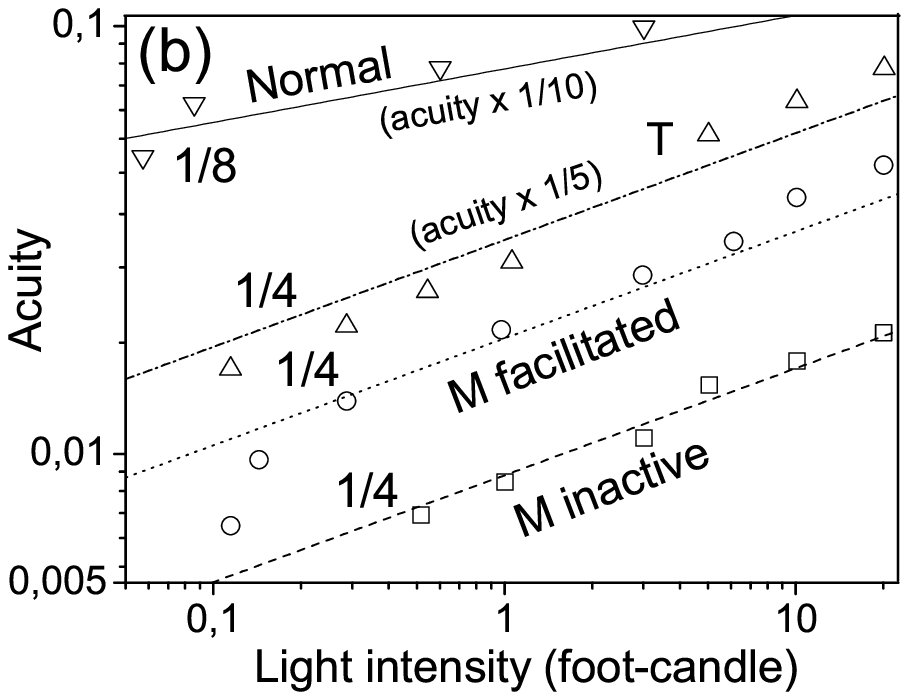}\\
\vspace*{0.6cm}
\includegraphics[width=5.5cm]{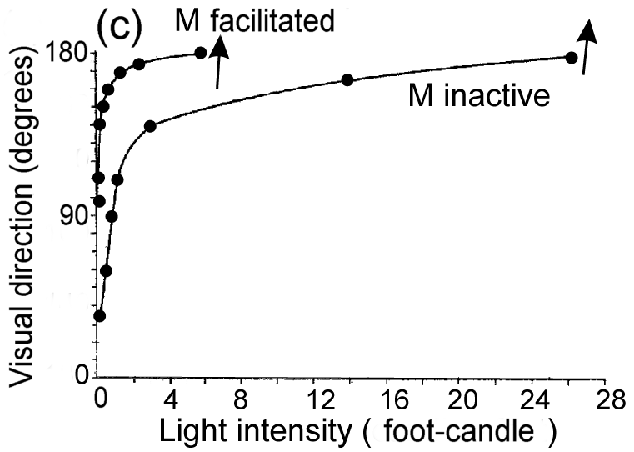}
\end{center}
\caption{\label{PLAWSTI} (a) Visual field of right eye versus
relative illumination (test object: 1 cm diameter, white disk).  M
($m \approx 1/4$), M facilitated ($m \approx 1/4$),  T ($m \approx
1/8$). (b) Acuity of right eye versus illumination. M ($m \approx
1/4$), M facilitated ($m \approx 1/4$), T ($m \approx 1/4$) and
normal man ($m \approx 1/8$). (Adapted from Fig. 1(a)(b) of
\cite{Gonzalo07P} with permission of Springer Science and Business
Media). (c) Visual direction function versus illumination of an
upright test arrow. (Adapted from Fog. 3(c) of \cite{Gonzalo96}
with permission of The MIT Press).}
\end{figure}

Concerning the effect of the intensity of the stimulus, Fig.
\ref{PLAWSTI} shows some examples of visual functions or qualities
versus intensity of stimulation in a stationary regime
\cite{Gonzalo45,Gonzalo52}. The well-known Stevens law relating
perception $P$ and stimulus $S$ by a power law of the type
\begin{equation}
P= pS^m  \label{STEVENS}
\end{equation}
(considered an improvement of the Fechner law) was used to fit the
data, the slope of the straight line in a log-log representation
being the exponent $m$. Fig. \ref{PLAWSTI} (a) shows the visual
field amplitude of right eye in cases M, M facilitated by strong
muscular stress (40 kg held in his hands), and case T, versus the
illumination of a test object. As seen, the data fit rather well
to Stevens straight lines, not very close of the highest values.
The slope $m$ of the fitting straight lines is remarkably close to
1/4 for M and M facilitated, and 1/8 for T. In Fig.
\ref{PLAWSTI}(b), similar representation is shown for the visual
acuity in central vision, including a normal case. Straight lines
with slope 1/4 fit rather well to the data of case M inactive, the
central range of the data for M facilitated and T; and with slope
1/8 for a normal case. In Fig. \ref{PLAWSTI}(c), it is shown in
linear scale representation, how the perceived direction of an
upright test white arrow  tends to the normal value (upright
direction) as the illumination of the arrow increases, for M
inactive and facilitated. These data do not show an admissible
agreement with a simple power law and the curves merely join the
experimental points.

\begin{figure}
\begin{center}
\includegraphics[width=5.5cm]{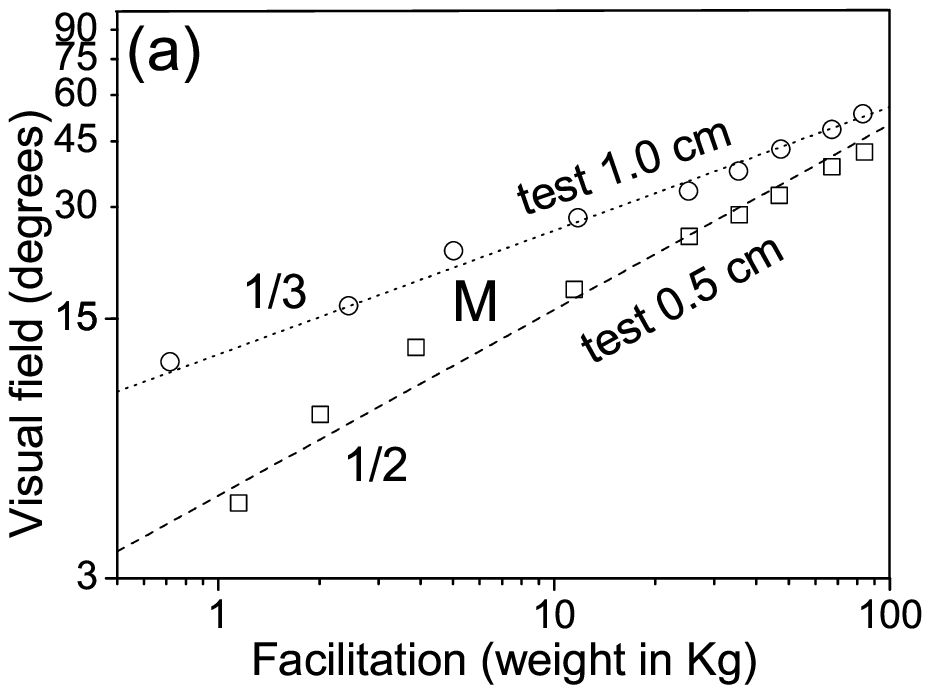}\\
\vspace*{0.3cm}
\includegraphics[width=5.5cm]{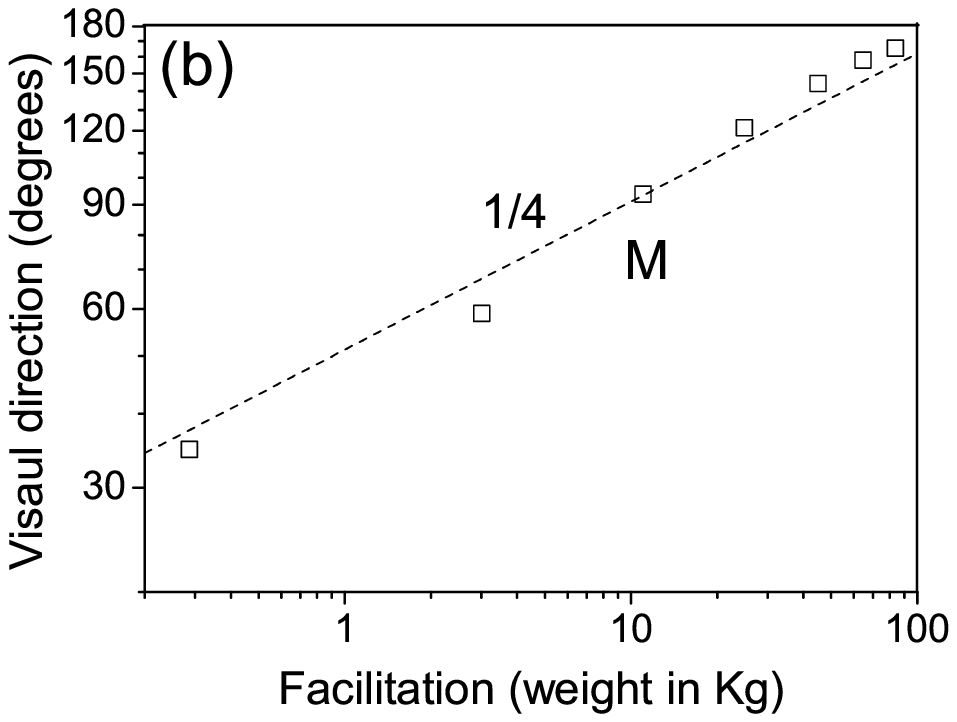}\\
\vspace*{0.3cm}
\includegraphics[width=5.5cm]{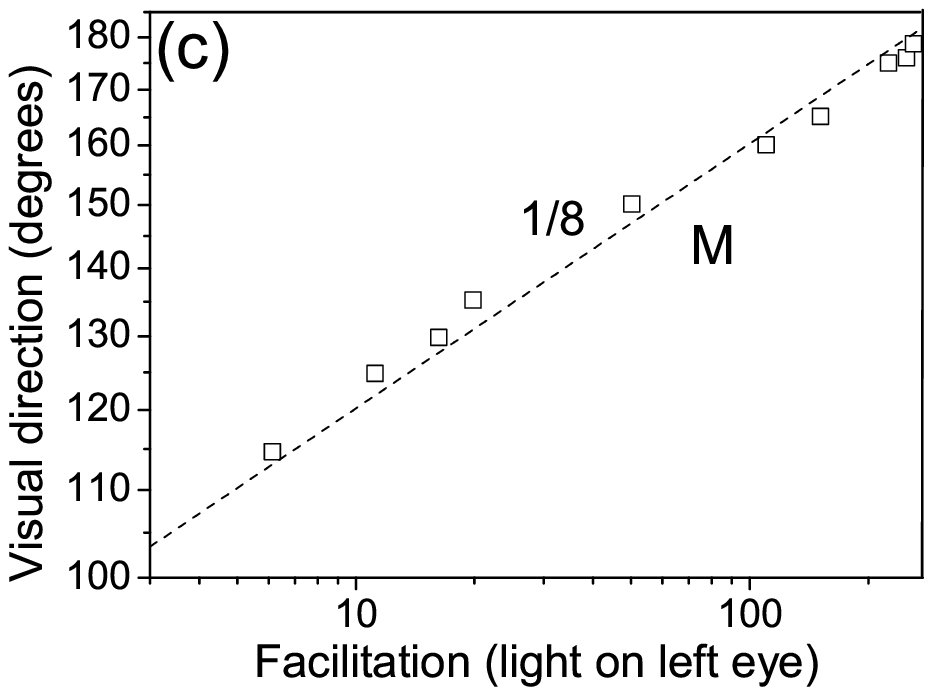}
\end{center}
\caption{\label{PLAWFACIL} Case M, right eye. (a) Log-log
representation of visual field amplitude versus facilitation by
muscular stress holding weight in his hands. Squares: 0.5 cm
diameter test size ($m \approx 1/2$). Circles: 1.0 cm diameter
test size ($m \approx 1/3$). (b) Visual direction (reinversion)
versus facilitation by muscular stress ($m\approx 1/4$). (c)
Visual direction versus facilitation by illumination on left eye
($m\approx 1/8$). (From the respective Figs. 2(a), 2(b) and 3 of
\cite{Gonzalo07P}, with permission of Springer Science and
Business Media).}
\end{figure}

Concerning facilitation, its effect is shown in Fig.
\ref{PLAWFACIL}, in log-log representation. Now the intensity of
light on the test object was kept constant and low, say $s$. The
data correspond also to a stationary regime
\cite{Gonzalo45,Gonzalo52}. Fig. \ref{PLAWFACIL}(a) shows the
visual field amplitude of right eye in case M, versus facilitation
by muscular stress holding in his hands increasing weights. The
data fit  to straight lines with slopes (Stevens' powers) $1/2$
and $1/3$ for the two different diameters of the white circular
test object. A greater test object implies a greater stimulus, and
 we can see that it leads to a lower slope, i.e., to a lower effect
of the facilitation. Fig. \ref{PLAWFACIL} (b) shows data under
similar conditions as in Fig. \ref{PLAWFACIL} (a) but the sensory
function measured is the recovery of the upright direction
($180$º) of an upright test white arrow that the patient perceived
tilted or almost inverted ($0$º) under low illumination. The data
fit to power $1/4$. The novelty in Fig. \ref{PLAWFACIL}(c) is that
facilitation is supplied by illuminating the left eye  which is
not observing the object, the power of the fitting being $1/8$
\cite{Gonzalo07P}. Note that the fittings of facilitation are much
better than those of simple stimulation.

The sensory growth by facilitation through other stimuli is a
multisensory or cross-modal effect in the cerebral integrative
process. Along all the cases reported in the literature on
reversal of vision \cite{Gonzalo07}, there are some of them in
which the image was reinverted by multisensory facilitation (e.g.,
\cite{Arjona02, River98}). Facilitation by multisensory or
cross-modal effects have been also observed in patients with
visual deficits, e.g. \cite{Frassinetti05,Poggel06}, and also in
normals,
e.g.\cite{Martuzzi07,Diederich07,Gillmeister07,Holmes07,Rowland07,Baier06,Schnupp05}.

The capability to improve the perception in central syndrome by
multisensory facilitation was found to be greater as the deficit
of excitability in the reduced cerebral system is greater, and as
the primary stimulus (e.g., the constant value of light $s$ and
the size of the test object) is weaker \cite{Gonzalo45} (see Fig.
\ref{PLAWFACIL}(a)). The first of these conditions is in relation
to what is observed recently
\cite{Laurienti06,Peiffer07,Poggel06}, and the last condition is
also in agreement with recent observations
\cite{Gillmeister07,Schnupp05}.It was suggested \cite{Gonzalo03}
that the multisensory interaction in central syndrome leads to a
nonlinear contribution to the cerebral excitation with stimuli,
this being also in relation to recent suggestions
\cite{Standford07}.

For the Stevens power laws,  note that their validity is assumed
to be restricted to  limited ranges of stimuli, and they are not
exempt from criticism. However, in connection with these laws, it
is remarkable that many biological observable quantities are
statistically seen to scale with the {\em mass} $M$ of the
organism, according to power laws $Y= k M^r$, where most of the
exponents $r$ are multiples of 1/4. These scaling laws are
supposed to arise from universal mechanisms in all biological
systems (in our case, neural networks) as the optimization to
regulate the activity of its subunits, as cells
\cite{West05,Anderson01}. Under the assumption that a stimulus $S$
activates a neural mass $M_{neur}= \alpha S^{\beta}$
\cite{Arthurs04,Gonzalo07P}, the above power law of perception
(\ref{STEVENS}) becomes $P= k (M_{neur})^{\beta /m}= k
(M_{neur})^r$, i.e., we recover a biological scaling power law of
growth which would be the basis of the perception laws shown
above. In cases where $\beta$ would be close to unity
\cite{Arthurs04}, then $r\approx m$, and Stevens laws would
exhibit quarter powers as seen in some of the cases here analyzed.

\section{Conclusions}

The functional model described highlights a functional unity and
continuity of the cerebral cortex. This is reflected in the
functional gradients according to which the specificity of the
cortex is distributed with a continuous variation,  giving place
to regions where several specificities overlap significantly. A
functional unity is suggested in the sense  that for a sensory
function to be normal, the integrative process must involve the
whole specific gradient extended over the cortex and not only over
the region with higher functional density. Multisensory
integration would be involved in a greater degree in regions where
the overlapping of the specific functional gradients is greater.
This scheme affords an interpretation of the great variety of
syndromes and responds to requirements formulated recently, as
said in the introduction.

A neural mass lost in the rather unspecific region where the
decline of several gradients overlap, as in the so-called central
zone, is interpreted as a scale reduction in the nervous
excitability of the cerebral system. The functional gradients
associated to the different sensory qualities of a given sensory
system  become then affected allometricaly, leading to stages in
perception  of incomplete integration in a different degree for
each quality. Inverted or tilted perception appears as an stage of
incomplete integration. The  higher or complex functions, which
require greater nervous excitation (and then integration), are
more affected  and  lost first. A functional continuity is
manifested from elementary sensory functions to  higher ones,
according to their excitability thresholds, following a
physiological order, as is shown in the dynamic disgregation of
qualities as excitation diminishes. Allometric scaling laws
account for the continuous variation of each quality (e.g., visual
direction, acuity) in relation to another quality (e.g., visual
field).

A growth in the sensory level, and then a more complete
integration, is achieved by means of an increase in the primary
stimulus or by multisensory facilitation, mechanisms which supply
in part the neural mass lost. This capability is greater as the
deficit is greater and the primary stimulus is weaker. The growth
of the sensory level follows approximately scaling power laws that
would emerge from the dynamics of biological neural networks.

Note that the approach exposed could be connected with those based
on a distributed character of the cerebral processing and its
adaptive and long-distance integrative aspects (e.g,
\cite{Yuval07,Rodriguez99}).

\vspace{0.3cm}

Acknowledgement. I am very grateful to M.A. Porras for his
valuable assistance in preparing this manuscript.


\begin{thebibliography}{99}

\bibitem{Martuzzi07} R. Martuzzi, M.M.Murray, C.M. Michel et al.,
Multisensory interactions within human primary cortices revealed
by BOLD dynamics, Cerb. Cortex 17 (2007) 1672-1679.

\bibitem{Gillmeister07} H. Gillmeister  and M. Eimer, Tactile
enhancement of auditory detection and perceived loudness, Brain
Res. 1160 (2007) 58-68.



\bibitem{Kayser07} C. Kayser and N.K. Logothetis, Do early sensory cortices
integrate cross-modal information?, Brain Struct. Funct. 212
(2007) 121-132.

\bibitem{Kayser072} C. Kayser, C.I. Petkov, M. Augath and N.K.
Logothetis, Functional Imaging reveals visual modulation of
specific fields in auditory cortex, J. Neurosci. 27 (2007)
1824-1835.


\bibitem{Alvarado07} J.C. Alvarado, J.W. Vaughan, T.R. Stanford and B.E. Stein, Multisensory versus
Unisensory Integration: Contrasting Modes in the Superior
Colliculus, J. Neurophysiol.  27 (2007) 1824-1835.

\bibitem{Diederich07} A. Diederich  and H. Colonius, Why two
``Distractors" are better than one: modeling the effect of
non-target auditory and tactile stimuli on visual saccadic
reaction time, Ex. Brain Res. 179(2007) 43-54.


\bibitem{Poggel06} D.A. Poggel, E. Kasten, E.M. Muller-Oehring et
al., Improving residual vision by attentional cueing in patients
with brain lesions, Brain Res. 1097 (2006) 142-148.

\bibitem{Bizley06} J.K. Bizley, F.R. Nodal, V.M. Bajo  et al.,
Physiological and Anatomical Evidence for Multisensory
Interactions in Auditory Cortex, Cereb. Cortex 17 (2007)2172-2189.

\bibitem{Frassinetti05} F. Frassinetti, N. Bolognini, D. Bottari
et al., Audiovisual integration in patients with visual deficit,
J. Cogn. Neurosci. 17 (2005) 1442-1452.

\bibitem{Macaluso05} E. Macaluso, C.D.
Frith, J. Driver, Multisensory stimulation with or without
saccades: fMRI evidence for crossmodal effects on sensory-specific
cortices that reflect multisensory location-congruence rather than
task-relevance, Meuroimage 26 (2005) 414-425.


\bibitem{Pascual04} H. Théoret, L. Merabet, A. Pascual-Leone, Behavioral and neuroplastic changes in
the blind: evidence for functionally relevant cross-modal
interactions, J. Physiol. Paris. 98 (2004) 221-33.



\bibitem{Calvert04} G.A. Calvert and T. Thesen, Multisensory
integration: methodological approaches and emerging principles in
the human brain, J. Physiol. Paris 98 (2004) 191-205.


\bibitem{Sathian02} K. Sathian and A.Zangaladze,  Feeling with the
mind´s eye: contribution of visual cortex to tactile perception,
Behav. Brain Res. 135 (2002) 127-132.

\bibitem{Pascual97} L.G. Cohen, P. Celnik, A. Pascual-Leone et al.,
% B. Corwell, L. Falz, J. Dambrosia, M. Honda, N. Sadato, C. Gerloff, M.D. Catalá, M. Hallett,
Functional relevance of cross-modal plasticity in blind humans,
Nature 389 (1997) 180-3.


\bibitem{Wallace04} M.T. Wallace, R. Ramachandran and B.E. Stein, A revise view of sensory cortical
parcellation, Proc. Natl. Acad. Sci. USA 101 (2004) 2167-2172.

\bibitem{Gonzalo45} J. Gonzalo, Investigaciones sobre la nueva din\'amica cerebral.
La actividad cerebral en funci\'on de las condiciones din\'amicas
de la excitabilidad nerviosa  (Publicaciones del Consejo Superior
de Investigaciones Cient\'{\i}ficas, Inst. S. Ram\'on y Cajal,
Madrid, Vol. I  1945,  Vol. II  1950). (Avalable in: Instituto
Cajal, CSIC, Madrid).

\bibitem{Gonzalo51} J. Gonzalo,  La cerebraci\'on sensorial y el desarrollo en espiral.
Cruzamientos, magnificaci\'on, morfog\'enesis, Trab. Inst. Cajal
Invest. Biol. 43 (1951) 209-260.

\bibitem{Gonzalo52} J. Gonzalo, Las funciones cerebrales humanas seg\'un nuevos datos y
bases fisiol\'ogicas: Una introducci\'on a los estudios de
Din\'amica Cerebral,  Trab. Inst.  Cajal Invest. Biol. 44 (1952)
195-157.

\bibitem{Goldstein18} K. Goldstein and A. Gelb, Psychologische Analysen hirnpathologischer
Fälle auf Grund Untersuchungen Hirnverletzer, Zeitschrift für die
gesamte Neurologie und Psychiatrie 41 (1918) 1-142.

\bibitem{Gonzalo96} I. Gonzalo and A. Gonzalo,   Functional gradients in cerebral dynamics:
The J. Gonzalo theories of the sensorial cortex, in: R.
Moreno-D\'{\i}az and J.Mira,  eds.,  Brain Processes, Theories and
Models. An Int. Conf. in honor of W.S. McCulloch 25 years after
his death (The MIT Press, Massachusetts 1996) 78-87.

\bibitem{Gonzalo97} I. Gonzalo,   Allometry in the J. Gonzalo's model of the sensorial
cortex,  Lect. Not.  Comp. Sci. 1240 (1997) 169-177.

\bibitem{Gonzalo99} I. Gonzalo,  Spatial Inversion and Facilitation in the J. Gonzalo's
Research of the Sensorial Cortex. Integrative Aspects,  Lect. Not.
Comp. Sci. 1606 (1999) 94-103.

\bibitem{Gonzalo01} I. Gonzalo and M.A. Porras, Time-dispersive effects in the J. Gonzalo's
research on cerebral dynamics,  Lect.  Not.  Comp. Sci.  2084
(2001) 150-157.

\bibitem{Gonzalo03} I. Gonzalo and M.A. Porras, Intersensorial summation as a nonlinear
contribution to cerebral excitation,  Lect.  Not.  Comp.  Sci.
2686 (2003) 94-101.

\bibitem{Gonzalo07} I. Gonzalo-Fonrodona, Inverted or tilted
inversion disorder, Rev. Neurol. 44 (2007) 157-165.

\bibitem{Gonzalo07P} I. Gonzalo-Fonrodona  and M.A. Porras,
Physiological Laws of Sensory Visual System in Relation to Scaling
Power Laws in Biological Neural Networks, Lect. Not. Comp. Sci.
4527 (2007) 96-102.

\bibitem{Critchley53} M Critchley, The Parietal Lobes (Arnold, London 1953) Chap 9.

\bibitem{Bender48} M.B. Bender and H.L.  Teuber, Neuro-ophthalmology,
Prog. Neurol.  Psychiatry III (1948) 163-182.

\bibitem{Ajuriaguerra49}   J.de Ajuriaguerra J and H. H\'ecaen, Le Cortex C\'er\'ebral \'Etude
Neuro-psycho-pathologique (Masson, Paris 1949).


\bibitem{Delgado78} A.E. Delgado, Modelos Neurocibern\'eticos de Din\'amica Cerebral.
Ph.D.Thesis,  E.T.S. de Ingenieros de Telecomunicaci\'on,
Universidad Polit\'ecnica de Madrid, 1978.

\bibitem{Mira87} J. Mira, A.E. Delgado and R. Moreno-D\'iaz, The fuzzy paradigm for
knowledge representation in cerebral dynamics,  Fuzzy Sets and
Systems 23 (1987) 315-330.

\bibitem{Mira95} J. Mira, A. Manjarr\'es, S. Ros et al.,
Cooperative Organization of Connectivity Patterns and Receptive
Fields in the Visual Pathway: Application to Adaptive Thresholdig,
 Lect. Not.  Comp. Sci. 930 (1995) 15-23.

\bibitem{Manjarres00} A. Manjarr\'es,  Modelado Computacional de la
Decisi\'on Cooperativa: Perspectivas Simb\'olica y Conexionista.
Ph.D. Thesis,  Ciencias  F\'{\i}sicas, Facultad de Ciencias de la
UNED, Madrid, 2001.


\bibitem{Arias04} M. Arias and I. Gonzalo, La obra neurocient\'{\i}fica de Justo Gonzalo
(1910-1986): El s\'{\i}ndrome central y la metamorfopsia
invertida, Neurolog\'{\i}a  19 (2004) 429-433.

\bibitem{Barraquer05} L. Barraquer,  La din\'amica cerebral de Justo Gonzalo en la
historia,  Neurolog\'{\i}a 20 (2005) 169-173.


\bibitem{River98} Y. River, T. Ben Hur and  I. Steiner, Reversal of vision
metamorphopsia, Arch. Neurol. 53 (1998)  1362-1368.

\bibitem{Arias01} M. Arias, C. Lema, I. Requena et al.,
Metamorfopsia invertida: una alteraci\'on en la percepci\'on de la
situaci\'on espacial de los objetos,  Neurolog\'{\i}a  16 (2001)
149-153.

\bibitem{Arjona02} A. Arjona and E. Fern\'andez-Romero,  Ilusi\'on de inclinaci\'on de
la imagen visual. Descripci\'on de dos casos y revisi\'on de la
terminolog\'{\i}a,  Neurolog\'{\i}a  17 (2002) 338-341.

\bibitem{Malis03} D.D. Malis and J.P. Guyot, Room tilt illusion as a manifestation of
peripheral vestibular disorders, Ann. Otol. Rhinol. Laryngol. 112
(2003) 600-605.

\bibitem{Hernandez06} A. H. Hern\'andez, F. Pujadas, F. Purroy et al.,
Upside down reversal of vision due to an isolated acute cerebellar
ischemic infarction, J. Neurol. 253 (2006) 953-954.

\bibitem{Unal06} A. Unal, A. Cila and S. Saygi, Reversal of vision
metamorphopsia: A manifestation of focal seizure due to cortical
dysplasia, Epilepsy Behav. 8 (2006) 308-311.

\bibitem{Kasten06} E. Kasten, and D.A. Poggel, A Mirror in the Mind: A Case of Visual
Allaesthesia in Homonymous Hemianopia, Neurocase 12 (2006) 98-106.

\bibitem{Perkkio85}J. Perkkiö and R. Keskinen, The relationship between growth and
allometry, J. theor. Biol. 113 (1985) 81-87.

\bibitem{Holmes07} N.P. Holmes, G.A. Calvert and C. Spence, Tool use
changes multisensory interactions in seconds: evidence from
crossmodal congruency task. Exp. Brain Res. 183 (2007) 465-476.

\bibitem{Rowland07} B.A. Rowland, S. Quessy, T.R. Standford and B.E. Stein,
Multisensory integration shortens physiological latencies, J.
Neurosc. 27 (2007) 5879-5884.

\bibitem{Baier06} B. Baier, A. Kleinschmidt and N.G. Müller, Cross-modal
processing in early visual and auditory cortices depends on
expected statistical relationship of multisensory information, J.
Neurosci. 26 (2006) 12260-12265.

\bibitem{Schnupp05} J.W. Schnupp, K.L. Dawe and G.L. Pollack, The
detection of multisensory stimuli in an orthogonal sensory space,
Exp. Brain Res. 162 (2005) 181-190.

\bibitem{Laurienti06} P.J. Laurienti, J.H. Burdette, J.A. Maldjian
and M.T. Wallace, Enhanced multisensory integration in older
adults, Neurobiol Aging 27 (2006) 1155-1163.

\bibitem{Peiffer07} A.M. Peiffer, J.L. Mozolic, C.E. Higenschmidt
and P.J. Laurienti, Age-related multisensory enhancement in a
simple audiovisual detection task, Neuroreport 18 (2007)
1077-1081.

\bibitem{Standford07} T.R. Standford and B.E. Stein, Superadditivity
in multisensory integration: putting the computation in context,
Neuroreport 18 (2007) 787-792.

\bibitem{West05} G.B. West and J.H. Brown, The origin of
allometric scaling laws in biology from genomes to ecosystems:
towards a quantitative unifying theory of biological structure and
organization, J. Exper. Biol.  208 (2005) 1575-1592.

\bibitem{Anderson01} R.B. Anderson, The power law as an emergent property, Mem.
Cogn. 29 (2001) 1061-1068.

\bibitem{Arthurs04} O.J. Arthurs,  C.M.E. Stephenson, K. Rice et
al., Dopaminergic effects on electrophysiological and functional
MRI measures of human cortical stimulus-response power laws,
NeuroImage  21 (2004) 540-546.

\bibitem{Yuval07} S. Yuval-Greenberg, L. Deouell, What You See Is
Not (Always) What You Hear: Induced Gamma Band Responses Reflect
Cross-Modal Interactions in Familiar Object Recognition, J.
Neurosci. 27 (2007) 1090-1096.

\bibitem{Rodriguez99} E. Rodr\'iguez, N. George, J.P. Lachaux et al., Perception´s shadow: long-distance
synchronization of human brain activity, Nature 397 (1999)
430-433.

\end{thebibliography}
\end{document}